\documentclass{article}

\usepackage[preprint]{neurips_2021}

\usepackage[utf8]{inputenc} 
\usepackage[T1]{fontenc}    
\usepackage{hyperref}       
\usepackage{url}            
\usepackage{booktabs}       
\usepackage{amsfonts}       
\usepackage{nicefrac}       
\usepackage{microtype}      
\usepackage{xcolor}         
\usepackage{graphicx}
\title{Neural Network Solver for Coherent Synchrotron Radiation Wakefield Calculations in Accelerator-based Charged Particle Beams}

\author{
  Auralee Edelen \\
  SLAC National Accelerator Laboratory\\
  {edelen@slac.stanford.edu} \\
  \And
  Christopher Mayes\\
  SLAC National Accelerator Laboratory\\
  {cmayes@stanford.edu} \\
}

\begin{document}

\maketitle{}

\begin{abstract}
  
  Particle accelerators support a wide array of scientific, industrial, and medical applications.  To meet the needs of these applications, accelerator physicists rely heavily on detailed  simulations of the complicated particle beam dynamics through the accelerator. One of the most computationally expensive and difficult-to-model effects is the impact of Coherent Synchrotron Radiation (CSR). As a beam travels through a curved trajectory (e.g. due to a bending magnet), it emits radiation that in turn interacts with the rest of the beam. At each step through the trajectory, the electromagnetic field introduced by CSR (called the CSR wakefield) needs to computed and used when calculating the updates to the positions and momenta of every particle in the beam. CSR is one of the major drivers of growth in the beam emittance, which is a key metric of beam quality that is critical in many applications. The CSR wakefield is very computationally intensive to compute with traditional electromagnetic solvers, and this is a major limitation in accurately simulating accelerators. Here, we demonstrate a new approach for the CSR wakefield computation using a neural network solver structured in a way that is readily generalizable to new setups. We validate its performance by adding it to a standard beam tracking test problem and show a ten-fold speedup along with high accuracy. 

\end{abstract}

\section{Introduction and Motivation}
Particle accelerators support a wide array of scientific, industrial, and medical applications. To meet the requirements of these applications, accelerator physicists rely heavily on detailed physics simulations of the particle beam dynamics through the accelerator, both for initial design of the accelerator and subsequent experiment planning (e.g. finding optimal accelerator settings for new experiments). One of the most important and difficult-to-model beam dynamics effects comes from the impact of Coherent Synchrotron Radiation (CSR) [1,2,3,4]. In simple terms, as a beam is transported through a curved trajectory, the particles in the beam emit radiation that in turn hits and adds momentum to other particles in the beam, as shown in Fig. 1. The CSR effect is a major driver of growth in the beam emittance, denoted $\varepsilon$, which is the overall beam size in position-momentum phase space and is a critical metric of beam quality in many applications. For example, $\varepsilon$ has a large impact on the quality of light produced by free electron lasers (FELs) for scientists to interrogate biological, chemical, and material samples [5]. Figure 1c shows an example of the impact CSR has on an electron beam in the Linac Coherent Light Source (LCLS) [7], a world-leading FEL user facility, where $\varepsilon$ is increased by a factor of 2 due to CSR. CSR also has a major effect on highly-compressed beams, such as those that will be generated by the FACET-II accelerator [6].

Simulations of accelerators are often conducted using ``particle tracking" codes, which track the positions ($x$, $y$, $z$) and momenta ($p_x$, $p_y$, $p_z$) of particles in the beam (collectively, the 6D position-momentum phase space). As the beam travels through the accelerator, the phase space is iteratively updated based on forces from accelerator components (e.g. magnets for steering and focusing, rf cavities for accelerating) and any internal effects that the beam has on itself (e.g. CSR, self-fields due to beam charge, etc). To account for the CSR effect, the electromagnetic field introduced by CSR (i.e. the ``CSR wakefield") needs to be computed at each tracking step. This is then used to calculate the induced change in momentum (called the CSR kick, or $K_{CSR}$) for every particle in the beam. The impact of CSR from all particles at all previous times in the beam trajectory must be taken into account, making this a very computationally intensive effect to simulate. A variety of methods are used for this [1], and most simplify the problem to reduce the computational complexity. The contribution from the beam charge density $\lambda$ along $z$ has the greatest impact on the beam, and as such the vast majority of simulation codes only include this 1D effect. 1D CSR is still computationally expensive to compute; for example, in cases we examined using the simulation code Bmad [8], inclusion of CSR slows down the simulation by a factor of 10. 

Here we demonstrate a new approach for speeding up the CSR wakefield calculation by replacing the electromagnetic solver with a neural network (NN). The NN solver is constructed in a very general way, making it readily extensible for general use in accelerator simulations. We validate the NN solver's performance on a standard CSR benchmark problem. It is 10$\times$ faster to execute and shows good agreement with traditional solvers. 



\begin{figure}
  \centering
  \includegraphics[width=0.99\textwidth]{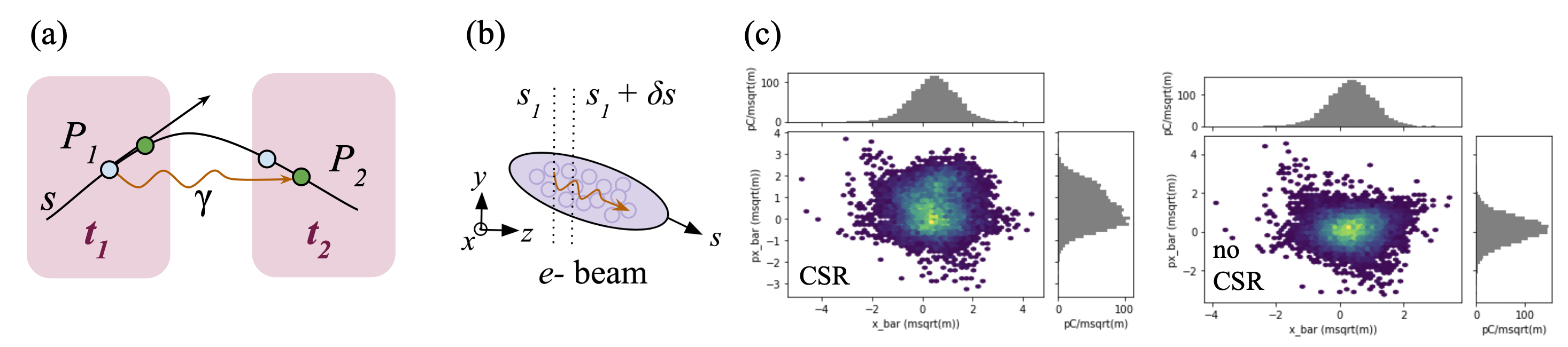}
  \caption{Simplified explanation of the CSR effect (a). A source particle (shown in blue) travels along curved trajectory $s$ some distance behind a second particle (shown in green). Radiation emitted from the source particle at time $t_1$  kicks (hits and adds momentum to) the second particle at $t_2$ after both have traveled further along $s$. The electromagnetic field due to CSR (the CSR wakefield) comes from contributions of the entire particle beam at all previous times of travel. (b) The CSR wakefield is calculated at each step through $s$ and is used to compute the change in each particle's position and momentum when advancing to the next tracking step $s+\delta s$. (c) An example of the impact of CSR for the LCLS accelerator, showing the transverse beam phase space ($x$ vs. $p_x$) with and without CSR.
  }
\end{figure}

  

\begin{figure}[hb]
  \centering
  \includegraphics[width=0.94\textwidth]{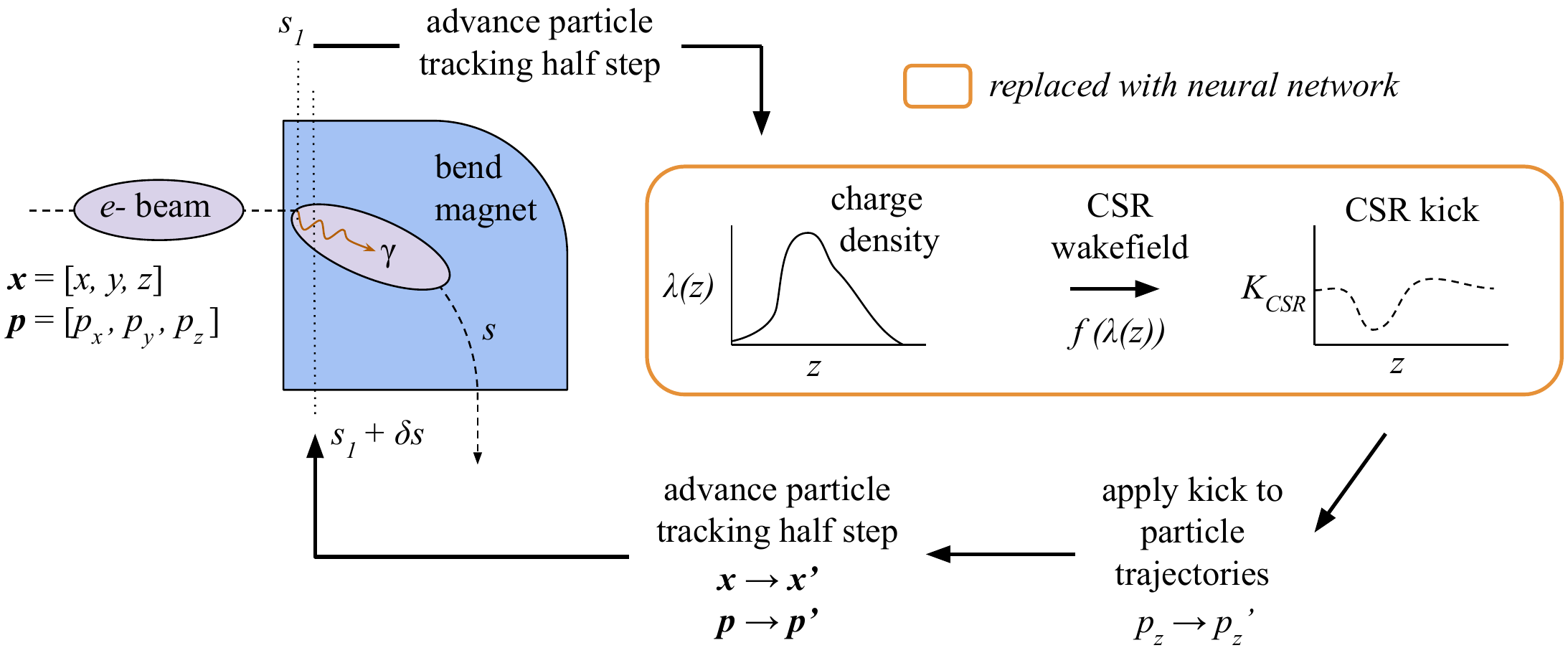}
  \caption{Particle tracking with 1D CSR. At step $s_1$, the particle beam phase space is advanced through half a standard tracking step to a new set of positions and momenta through a magnet. The charge density $\lambda$($z$) is then used to calculate the CSR wakefield. The wakefield is then used to compute the momentum kick $K_{CSR}$ that is imparted on every particle in the beam. Finally, the remaining half-step of standard tracking is completed. In this work, we replace the computationally expensive wakefield calculation with a NN that can predict $K_{CSR}$ from $\lambda$($z$) and $s$.}
\end{figure}

\begin{figure}[ht]
  \centering
  \includegraphics[width=0.45\textwidth]{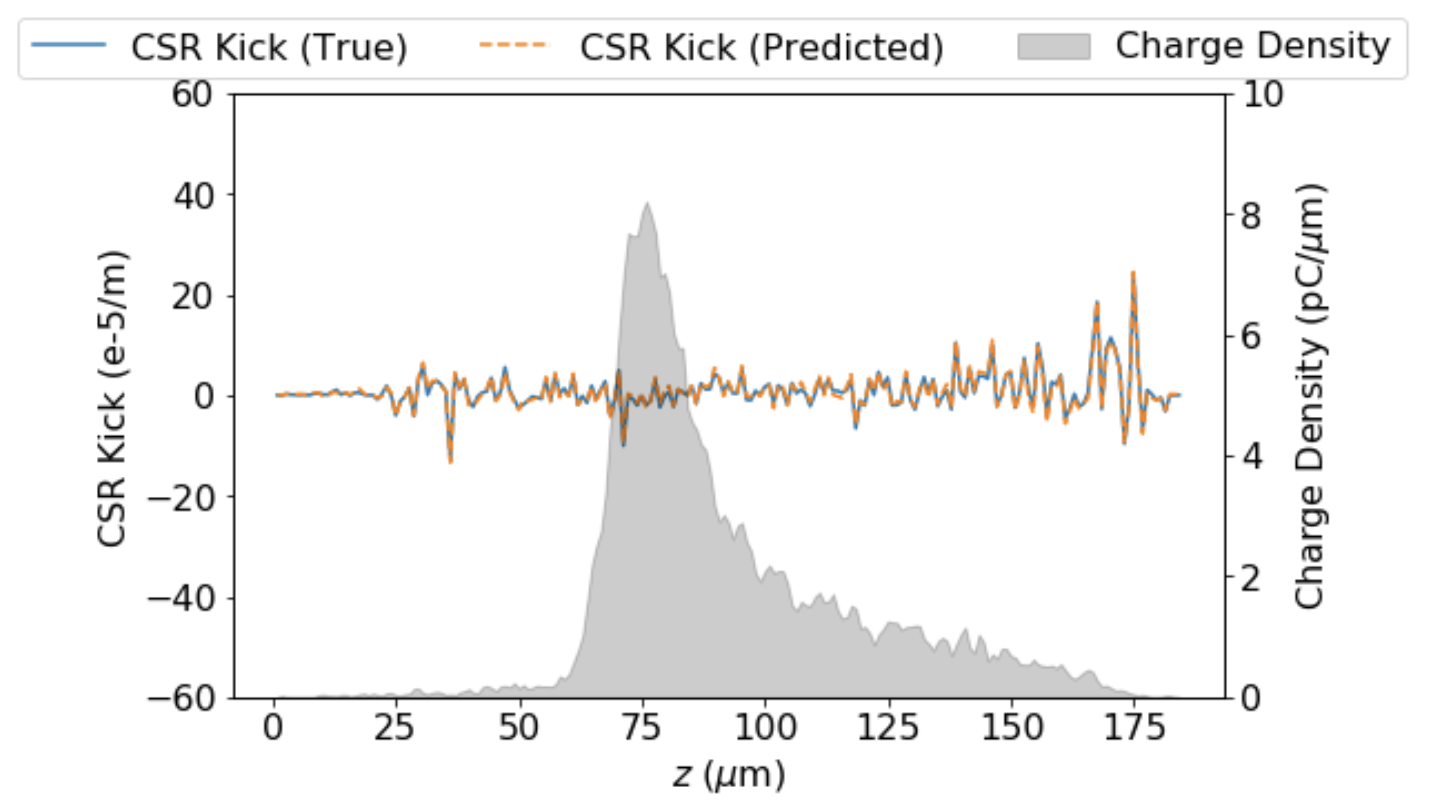}
    \includegraphics[width=0.45\textwidth]{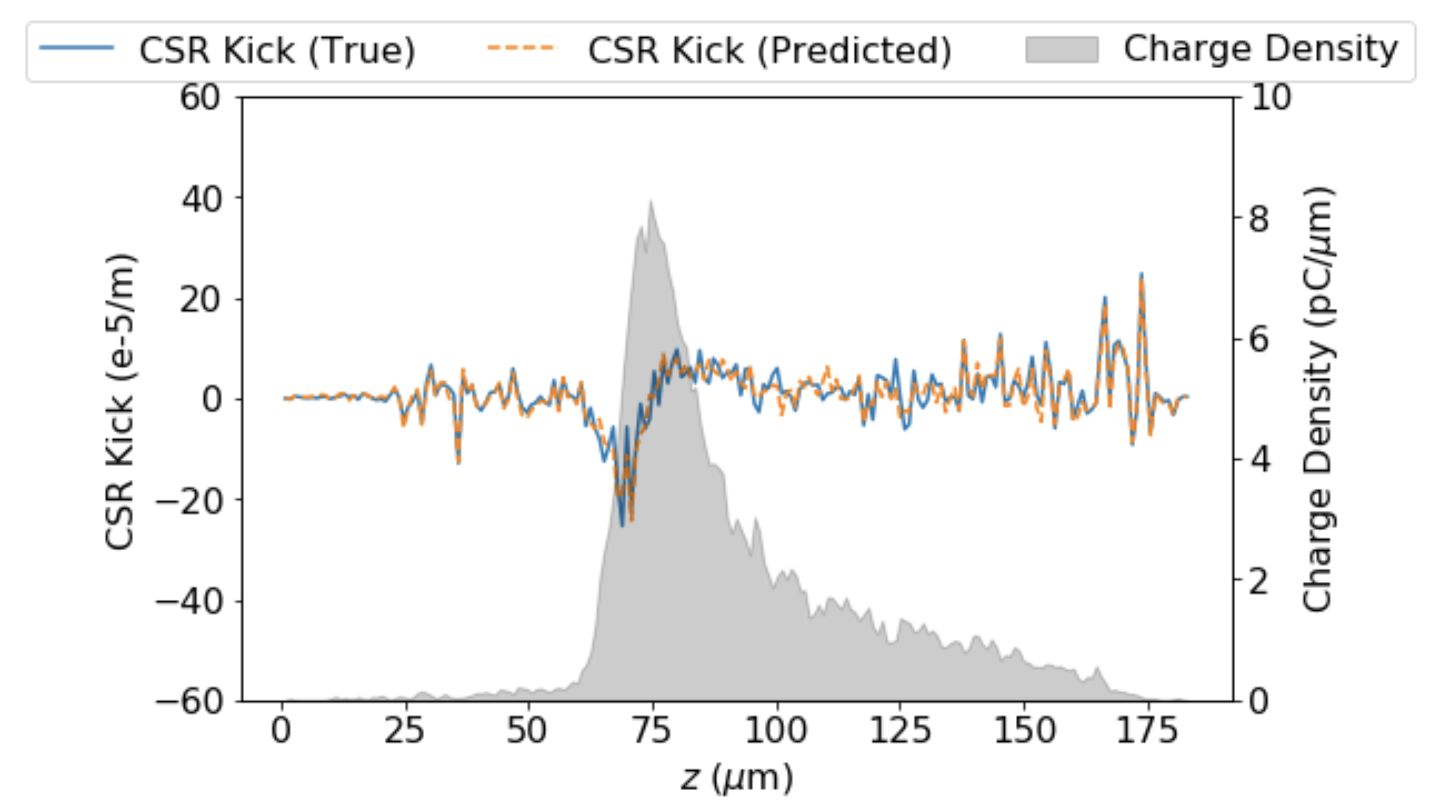}
    \includegraphics[width=0.45\textwidth]{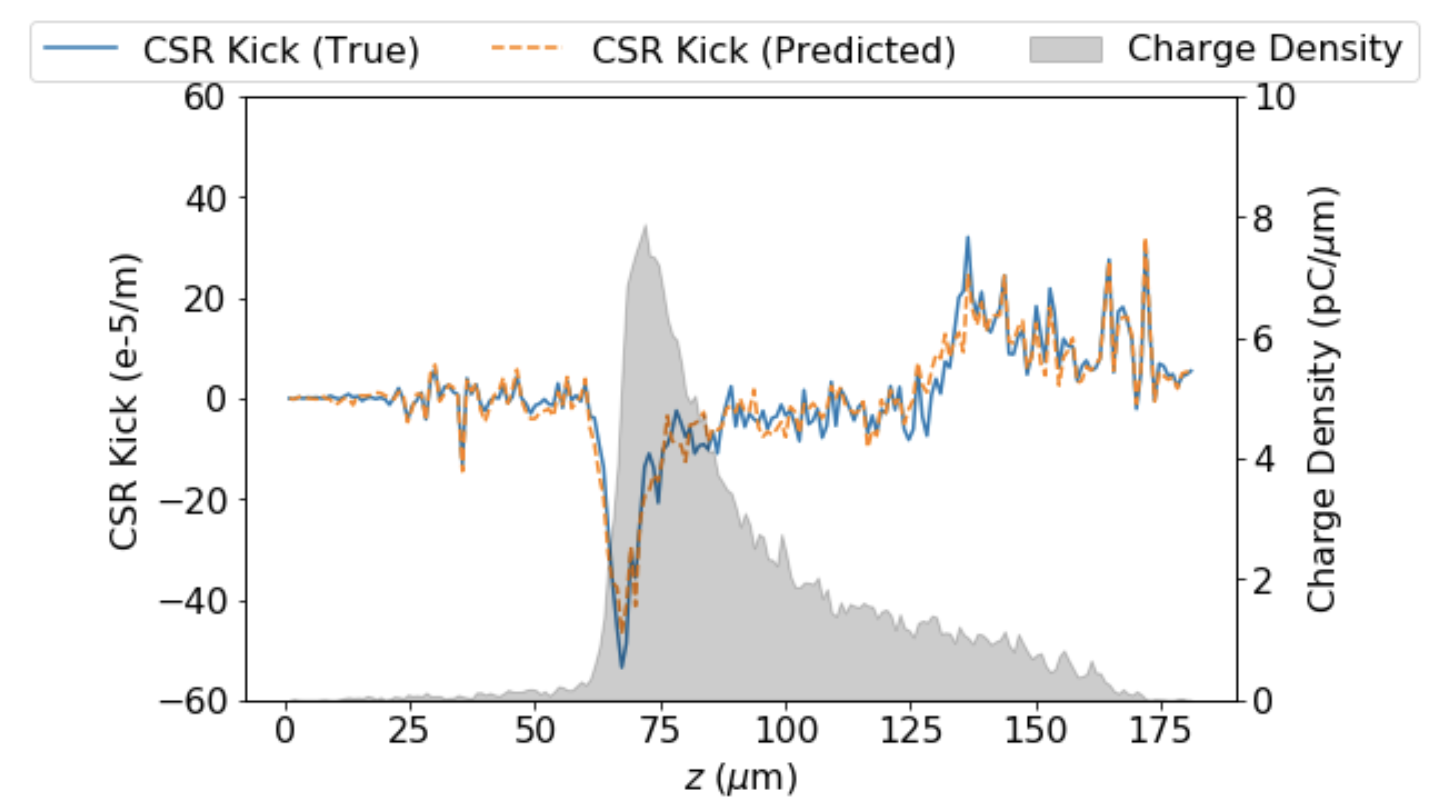}
    \includegraphics[width=0.45\textwidth]{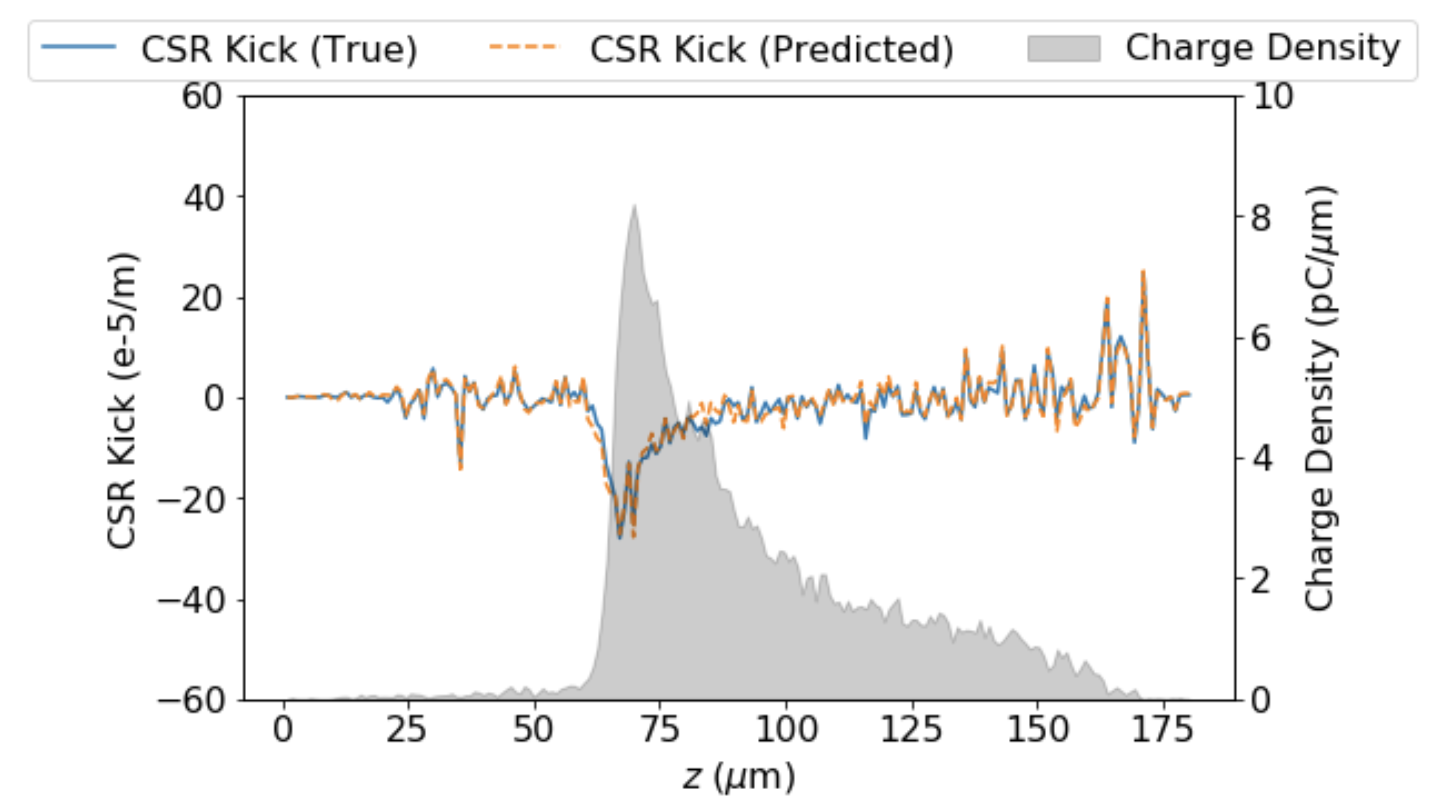}
  \caption{Examples of $K_{CSR}$ for a single beam calculated using a standard CSR solver at several different steps $s+\delta s$ into a magnet, compared with $K_{CSR}$ from the NN CSR solver.}
\end{figure}

\section{Problem Formulation}

In accelerator simulations, particles are ``tracked'' through accelerator components by updating their positions ($x$,$y$,$z$) and momenta ($p_x$,$p_y$, $p_z$) at each step $\delta s$ through the component. In the 1D formulation of the CSR problem, the computation at each step is as follows (see also  Fig. 2): 1. Conduct a half-step of standard particle tracking; 2. Calculate the charge density $\lambda$ as a function of $z$ in the beam; 3. Calculate $K_{CSR}$ from $\lambda$($z$) (for calculation details see [9]); 4. Compute the new $p_z$ for every particle based on  $K_{CSR}$; 5. Complete the second half-step the standard tracking.

Because the CSR effect is strongest in highly compressed (i.e. high density) beams, the last magnet in a beam compression chicane is often used as a standard benchmark problem for different ways of simulating CSR. Here we generated data from simulations of the second chicane in the LCLS [7], using a wide variety of realistic initial particle beams from upstream simulations. We used the beam dynamics code Bmad [9], which has a well-benchmarked 1D CSR solver [8,10]. Using this data, we trained a fully-connected, feed-forward NN to predict $K_{CSR}$ at each tracking step using the distance $s$ into the curved trajectory and $\lambda$($z$) as inputs. The original data for $\lambda$($z$) and $K_{CSR}$ were reduced into histograms of 200 bins, and the bunch length was used as an additional scalar input to account for the fixed bin width. A total of 1,234,000 samples were and split into training, validation, and testing sets in a 60-20-20 ratio. Additional test beams were also used to evaluate the NN performance (i.e. new beams with all tracking steps unseen). The NN has has 9 hidden layers with 200 nodes in each layer and $\tanh$ activation functions. It was trained using the Adam optimizer, and the learning rate was reduced by a factor of 0.8 on any detected plateau in the validation loss. Early stopping, an L2-norm penalty on the weights, and a dropout rate of 0.05 was used to prevent over-fitting. The code and the data can be found at [11]. All data were generated using 4 CPU nodes on Cori at NERSC [12], and all training was run on an NVIDIA GeForce 2080i GPU on an internal cluster. The NN construction and training was done in tensorflow [13] and keras [14].

\section{Neural Network CSR Solver Performance}

The first assessment of the NN solver is whether it accurately predicts $K_{CSR}$ for new test beams at each individual step $s+\delta s$. We find excellent agreement between the NN solver and Bmad's 1D CSR routine, with a mean absolute percent error of 2.8 between the predicted and true values. Fig. 3 shows the predicted $K_{CSR}$ for different time steps through the magnet for one example beam, and a video of the evolution can be viewed at [11].
The next assessment considers the compounding errors that are introduced when using the NN solver for each step in tracking. We assessed this by adding the NN solver to a simple Python-based beam tracking code and comparing the results with those obtained by Bmad. We also compare against the case where no CSR effects are included in the simple tracking code. We find that the NN solver reliably reproduces the expected effect on $\varepsilon$ at the end of tracking. The accuracy is especially high when compared with the impact of excluding CSR effects altogether. An example comparing the evolution of $\varepsilon$ through a magnet along $s$ is shown in Fig. 4, along with the final transverse beam distributions. In that example, $\varepsilon$ predicted with Bmad and the NN solver agree (1.4 mm-mrad), whereas the case that excludes CSR has a factor of 3 error (0.5 mm-mrad). Statistics for the performance on all test beams are also shown in Fig. 4. 

The major advantage of using the NN CSR solver is the computation speed that it attains while still accurately capturing the impact of CSR. Based on statistics over multiple runs for this benchmark problem, the traditional CSR computation in Bmad is about 10$\times$ slower than the computation without CSR. In contrast, the NN inference time is negligible and effectively restores the 10$\times$ faster execution.

\begin{figure}[hb]
  \centering
  \includegraphics[width=0.5\textwidth]{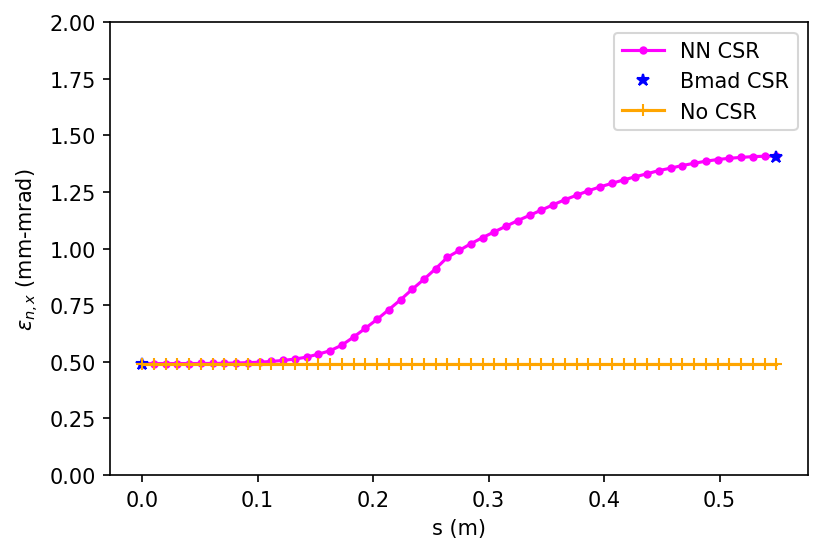}
   \includegraphics[width=0.36\textwidth]{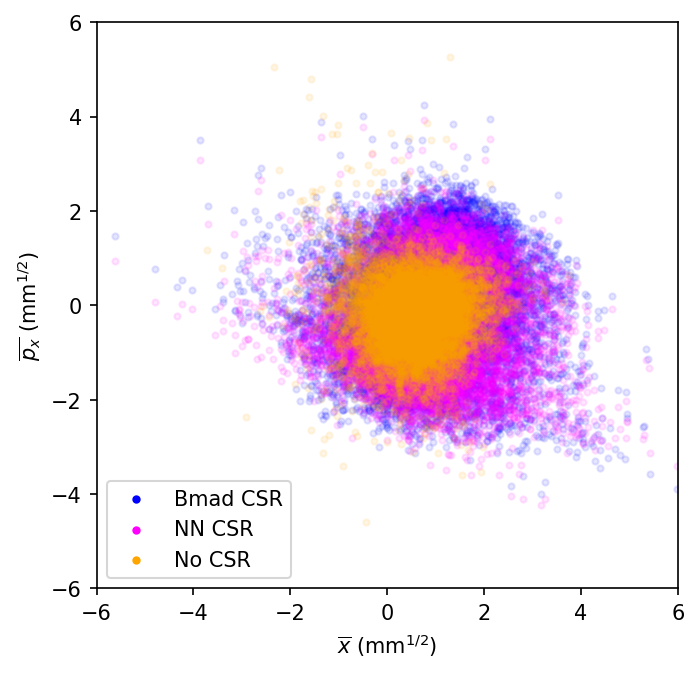}
   \\
  \includegraphics[width=0.36\textwidth]{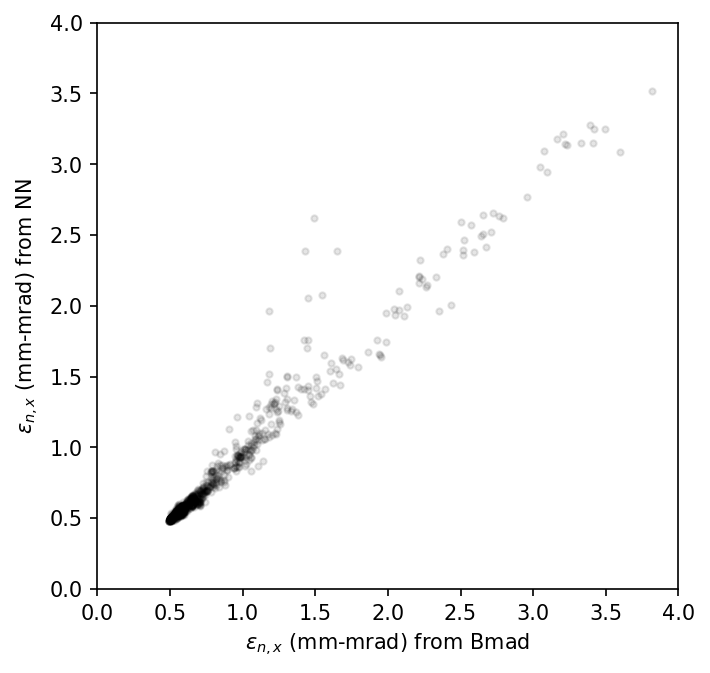}
  \includegraphics[width=0.5\textwidth]{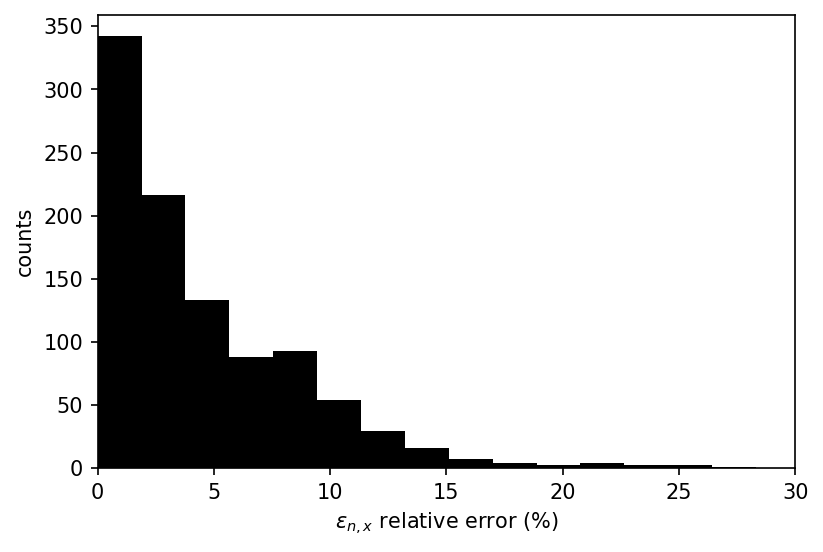}
  \caption{For an example beam, comparison of emittance evolution along $s$ with the NN CSR solver, no CSR, and 1D CSR in Bmad are shown (top left), along with the differences in the final $x$ vs. $p_x$ phase space in normalized coordinates (top right).  The summary statistics (bottom) for the comparison between the NN CSR solver and Bmad for 1k test beams show good agreement. The few outliers have smaller error than the exclusion of CSR would impart.
  }
\end{figure}


\section{Conclusions}

We demonstrated a new approach for speeding up the computation of the CSR effect in particle accelerator simulations using a NN solver. We validated the solver using a standard benchmark problem for 1D CSR and a wide variety of realistic particle beams. The NN solver showed good agreement with a standard 1D CSR solver used in Bmad and had a 10$\times$ faster execution speed. It is also substantially more accurate (e.g. 2--3 times) than excluding CSR in standard tracking. This makes it very appealing for cases where one wants to simulate the bulk effect of CSR quickly in design or optimization studies. The solver is set up in a general way, making it suitable to incorporate into general tracking codes, but this needs to be further validated on a wider set of problems. Some next steps include assessing how well it generalizes in practice to very different beams (e.g. much higher overall charge) and different magnet designs (e.g.  by including bending radius $\rho$ as an input parameter to the solver). We also plan to extend this to 2D CSR, but this presents additional challenges due to the computational expense of generating training data for that case (e.g. see [4]).


\section*{References}

\medskip

{
\small

[1] T. Agoh, ``Steady fields of coherent synchrotron radiation in a rectangular pipe'', Phys. Rev. ST Accel. Beams, Vol. 12, 094402, (2009). \url{https://journals.aps.org/prab/abstract/10.1103/PhysRevSTAB.12.094402}.

[2] G. Stupakov and S. Heifets, ``Beam instability and microbunching due to coherent synchrotron radiation'', Phys. Rev. Accel. Beams, Vol. 5, 054402, (2002). \url{https://journals.aps.org/prab/pdf/10.1103/PhysRevSTAB.5.054402}.

[3] E.L. Saldin, E.A. Schneidmiller, M.V. Yurkov, ``Radiative interaction of electrons in a bunch moving in an undulator'', Nuclear Instruments and Methods in Physics Research Section A, Vol. 417, Iss. 1, (1998).  \url{https://doi.org/10.1016/S0168-9002(98)00623-8}.

[4] W. Lou, C. Mayes, Y. Cai and G. White, ``Simulating two dimensional coherent synchrotron radiation
in python'', in Proceedings of 12th International Particle Accelerator Conference, Campinas, Brazil,
paper WEPAB234, (2021). \url{https://epaper.kek.jp/ipac2021/papers/wepab234.pdf}.

[5] M. Borland et al., ``Start-to-end jitter simulations of the linac coherent light source," Proceedings of the 2001 Particle Accelerator Conference, (2001). \url{https://ieeexplore.ieee.org/document/987880}.

[6] V. Yakimenko, et al., ``FACET-II facility for advanced accelerator experimental tests'', Phys. Rev. Accel. Beams, Vol. 22, 101301, (2019). \url{https://journals.aps.org/prab/abstract/10.1103/PhysRevAccelBeams.22.101301}.

[7] P. Emma, R. Akre, J. Arthur, et al. ``First lasing and operation of an ångstrom-wavelength free-electron laser'', Nature Photon 4, 641–647, (2010). \url{https://doi.org/10.1038/nphoton.2010.176}.

[8] D. Sagan, ``Bmad: A relativistic charged particle simulation library'', Nuclear Instruments and Methods in Physics Research Section A, Vol. 558, Iss. 1, (2006). \url{https://doi.org/10.1016/j.nima.2005.11.001}.

[9] D. Sagan, C.E. Mayes, ``Coherent Synchrotron Radiation Simulations for Off-Axis Beams Using the Bmad Toolkit'', in Proc. 8th Int. Particle Accelerator Conf. (IPAC'17), Copenhagen, Denmark, paper THPAB076, (2017). \url{https://doi.org/10.18429/JACoW-IPAC2017-THPAB076}.

[10] D. Sagan, G. Hoffstaetter, C. Mayes, ``Extended one-dimensional method for coherent synchrotron radiation including shielding'', Phys. Rev. Accel. Beams, Vol. 12, 040703, (2009). \url{https://link.aps.org/doi/10.1103/PhysRevSTAB.12.040703}.

[11] [redacted repository for anonymized review].

[12] National Energy Resource Computing Center (NERSC). \url{https://docs.nersc.gov/systems/cori/}. 

[13] TensorFlow: Large-scale machine learning on heterogeneous systems, 2015. Software available from tensorflow.org.

[14] F. Chollet, et al., Keras (2015), retrieved from \url{https://github.com/fchollet/keras}.

[15] A. Edeen, N. Neveu, M. Frey, Y. Huber, C. Mayes, and A. Adelmann, ```Machine learning for orders of magnitude speedup in multiobjective optimization of particle accelerator systems'', Phys. Rev. Accel. Beams, Vol. 24, 044601, (2020). \url{https://link.aps.org/doi/10.1103/PhysRevAccelBeams.23.044601}




}

\end{document}